\documentclass[fleqn,10pt]{wlscirep}
\usepackage[utf8]{inputenc}
\usepackage[T1]{fontenc}
\usepackage{algorithmic}
\usepackage[ruled]{algorithm2e}
\usepackage{bm}
\usepackage{chngcntr}
\title{Towards Computation- and Communication-efficient Computational Pathology}

\author[1,2,*,$\dagger$]{Chu Han}
\author[1,2,$\dagger$]{Bingchao Zhao}
\author[1,2,$\dagger$]{Jiatai Lin}
\author[2,3]{Shanshan Lyu}
\author[1,2]{Longfei Wang}
\author[1,2]{Tianpeng Deng}
\author[1,2]{Cheng Lu}
\author[1,2]{Changhong Liang}
\author[5]{Hannah Y. Wen}
\author[4,*]{Xiaojing Guo}
\author[1,2,*]{Zhenwei Shi}
\author[1,2,*]{Zaiyi Liu}

\affil[1]{Guangdong Provincial Key Laboratory of Artificial Intelligence in Medical Image Analysis and Application, Guangdong Provincial People's Hospital (Guangdong Academy of Medical Sciences), Southern Medical University, Guangzhou 510080, China}
\affil[2]{Department of Radiology, Guangdong Provincial People's Hospital (Guangdong Academy of Medical Sciences), Southern Medical University, Guangzhou 510080, China}
\affil[3]{Department of Pathology, Guangdong Provincial People's Hospital (Guangdong Academy of Medical Sciences), Southern Medical University, Guangzhou 510080, China}
\affil[4]{Department of Breast Pathology and Laboratory, Tianjin Medical University Cancer Institute and Hospital, National Clinical Research Center for Cancer, Key Laboratory of Breast Cancer Prevention and Therapy of Ministry of Education of China, Tianjin Medical University, Tianjin's Clinical Research Center for Cancer, West Huanhu Road, Tianjin, China, 300060}
\affil[5]{Department of Pathology and Laboratory Medicine, Memorial Sloan Kettering Cancer Center, 1275 York Avenue, New York, NY 10065}

\affil[*]{Correspondence: hanchu@gdph.org.cn (C.H.), guoxiaojing@tjmuch.com (X.G), shizhenwei@gdph.org.cn (Z.S.), liuzaiyi@gdph.org.cn (Z.L.)}
\affil[$\dagger$]{These authors contributed equally to this work.}

\begin{abstract}
Despite the impressive performance across a wide range of applications, current computational pathology (CPath) models face significant diagnostic efficiency challenges due to their reliance on high-magnification whole-slide image analysis. This limitation severely compromises their clinical utility, especially in time-sensitive diagnostic scenarios and situations requiring efficient data transfer. To address these issues, we present a novel computation- and communication-efficient framework called \underline{\textbf{MAG}}nification-Aligned \underline{\textbf{G}}lobal-\underline{\textbf{L}}ocal \underline{\textbf{Trans}}former (MAG-GLTrans). Our approach significantly reduces computational time, file transfer requirements, and storage overhead by enabling effective analysis using low-magnification inputs rather than high-magnification ones. The key innovation lies in our proposed magnification alignment (MAG) mechanism, which employs self-supervised learning to bridge the information gap between low and high magnification levels by effectively aligning their feature representations. Through extensive evaluation across various fundamental CPath tasks, MAG-GLTrans demonstrates state-of-the-art classification performance while achieving remarkable efficiency gains: up to 10.7$\times$ reduction in computational time and over 20$\times$ reduction in file transfer and storage requirements. Furthermore, we highlight the versatility of our MAG framework through two significant extensions: (1) its applicability as a feature extractor to enhance the efficiency of any CPath architecture, and (2) its compatibility with existing foundation models and histopathology-specific encoders, enabling them to process low-magnification inputs with minimal information loss. These advancements position MAG-GLTrans as a particularly promising solution for time-sensitive applications, especially in the context of intraoperative frozen section diagnosis where both accuracy and efficiency are paramount.

\end{abstract}
\begin{document}

\flushbottom
\maketitle

\thispagestyle{empty}

\section*{Introduction}
\label{sec:Introduction}
Digital slide scanners can convert the histopathology sections into calculable digital contents (whole-slide images, WSIs) and thus derive the field of computational pathology (CPath)~\cite{Song2023,srinidhi2021deep}. In the past few years, CPath has witnessed tremendous progress with the breakthrough of deep learning~\cite{lecun2015deep} in different fundamental CPath applications, such as cancer screening~\cite{bilal2023development}, cancer diagnosis~\cite{bulten2022artificial,lu2021ai}, risk stratification~\cite{amgad2024population,jiang2024end} and mutation prediction~\cite{wagner2023transformer,fu2020pan}. The accumulated multi-center and large-scale datasets also make CPath methods more generalizable and trustworthy in real-world clinical scenarios~\cite{xu2024whole,Shmatko2022,huang2023visual,vaidya2024demographic}.
Besides the tremendous success in diverse CPath applications, existing studies also propose various novel algorithms to tackle the technical challenges~\cite{Song2023}. For instance, weakly-supervised learning models are introduced to waive large among of manual annotations~\cite{laleh2022benchmarking,lu2021data,han2022multi}. Graph-based models extract the spatial and contextual interpretable features to provide clinically relevant evidence for the prediction results~\cite{Lee2022,Pati2021}. Federated learning frameworks protect the patients' privacy without data sharing across different medical centers~\cite{ogier2023federated,saldanha2022swarm}.

\begin{figure}[htp]
	\centering
	\includegraphics[width=.93\linewidth]{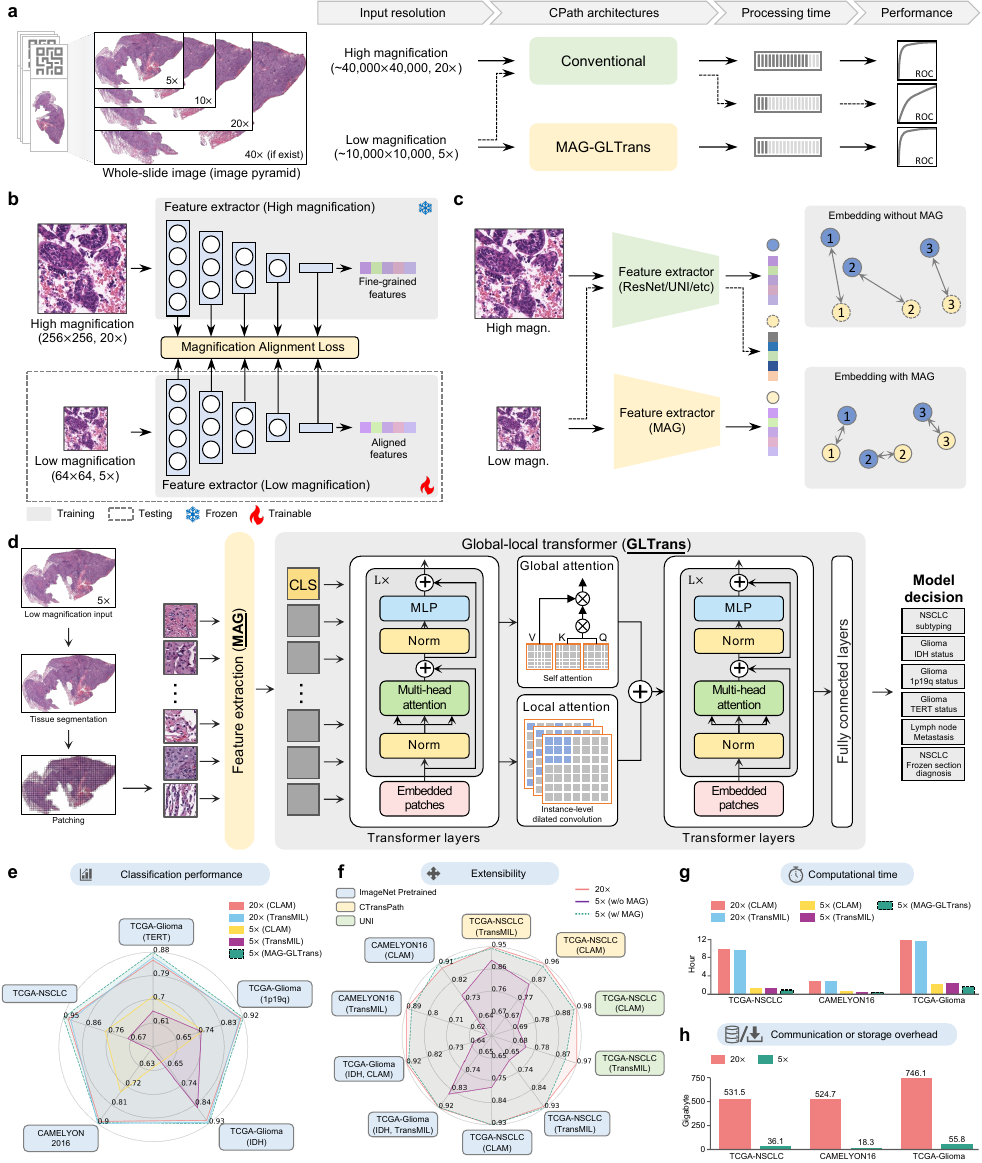}
	\caption{\textbf{Overview of the MAG-GLTrans framework.} \textbf{a}, Current computational pathology techniques analyze WSIs at high magnification, which poses significant diagnostic efficiency challenges. Our study introduces MAG-GLTrans, a computationally efficient solution that processes low-magnification input images to address these limitations. \textbf{b}, We propose magnification alignment (MAG) to preserve the information across different magnification levels in a self-supervised learning manner. \textbf{c}, MAG aligns the embedding spaces between high- and low-magnification images, enabling fine-grained feature extraction from low-resolution inputs. \textbf{d}, Model Architecture of MAG-GLTrans. It excels across various computational pathology (CPath) tasks in terms of classification performance, computational efficiency, communication efficiency, and extensibility. \textbf{e}, MAG-GLTrans outperforms two mainstream WSI analysis pipelines (CLAM~\cite{lu2021data} and TransMIL~\cite{shao2021transmil}) across five CPath tasks. \textbf{f}, MAG can be seamlessly extended to other WSI analysis architectures or histopathology-specific encoders (such as UNI~\cite{chen2024towards} or CTransPath~\cite{wang2022transformer}), enabling high performance even with low-magnification inputs. \textbf{g-h}, Processing the low magnification images can greatly save the computational time (\textbf{g}) as well as the communication and storage overhead (\textbf{h}).}
	\label{fig1}
\end{figure}
Nevertheless, the efficiency of WSI analysis pipelines is often overlooked in current studies, and this remains a significant challenge that greatly hinders their clinical viability. Typically, existing approaches~\cite{lu2021data,shao2021transmil} analyze WSIs under either the highest or second-highest magnification levels (40$\times$ or 20$\times$, Figure~\ref{fig1}a) to preserve fine-grained information~\cite{Song2023}. However, this scheme leads to critical efficiency issues. (1) Long computational time: Analyzing WSIs under high magnification levels will drastically increase the computational time due to the gigapixel resolution and the restriction of the computer hardware. It often takes tens of seconds or even minutes to analyze one WSI which may affect the diagnostic efficiency when facing a time-sensitive scenario such as intraoperative diagnosis~\cite{hollon2020near,vermeulen2023ultra}, or when dealing with large-scale WSIs. (2) High communication overhead: Computer-aided telemedicine can balance the inequality of the medical resources between urban and rural areas~\cite{marques2021enhanced,telemedicine98}. For telepathology, due to the huge image resolution, the immediacy of the computer-aided diagnostic algorithms is challenged and strongly depends on the network bandwidth, reliability of the connections across different sites and etc~\cite{SCHWEN2023100244}. In summary, to expand the universality of computational pathology methods in clinical practice, there is a need for a computation- and communication-efficient WSI analysis pipeline, while maintaining outstanding performance.

The major barrier to achieving the aforementioned goals is the gigapixel resolution of WSIs. The most intuitive solution to solve this challenge is to reduce the resolution of the images to be saved, transmitted and processed, which is a trade-off between efficiency and performance. Therefore, the major problem to be solved turns to how to bridge the information gap between the high resolution images and the low resolution ones. To achieve this, we have to resolve the following two arguments in both clinical and technical aspects. `\textit{Clinically, whether low resolution images contain diagnostic information?}' and `\textit{Technically, whether deep learning models can restore the missing information in the low resolution images?}' In clinical practice, pathologists can make an initial diagnosis at the low magnification. Existing studies also prove that macroscopic features can be used for some simple binary classification tasks~\cite{coudray2018classification,neary2023minimum}. In the technical aspect, deep learning-based models are possible to recover the information loss in various computer vision tasks, such as image/video super-resolution~\cite{dong2015image,kappeler2016video}, image colorization~\cite{zhang2016colorful} and image to video generation~\cite{zhu2020video}. The most advanced approach can even recover $64\times 64$ images to $1024\times 1024$ ones with up to 16 times recovery rate~\cite{saharia2022image}. It proves that deep learning models are capable of restoring the missing information and bridging the information gap between low and high resolution images. By resolving the above-mentioned two arguments, replacing high magnification inputs with low magnification ones for computation- and communication-efficient computational pathology is feasible while the research about this aspect is limited.

In this article, we propose MAG-GLTrans (Magnification-Aligned Global-Local Transformer), a computation- and communication-efficient framework designed for general-purpose WSI analysis. MAG-GLTrans dramatically accelerates computational time from minutes to just a few seconds compared to conventional computational pathology techniques (Figure~\ref{fig1}). Unlike traditional approaches that rely on high-resolution input images (20$\times$ or 40$\times$), MAG-GLTrans enables the use of lower-resolution inputs (10$\times$ or 5$\times$), as illustrated in Figure~\ref{fig1}d. By reducing the input dimension, MAG-GLTrans not only improve the computational efficiency but also minimizes communication overhead and significantly alleviates storage burdens. While lower image resolution typically implies information loss, MAG-GLTrans is capable of restoring information (Figure~\ref{fig1}b-c) by aligning the low magnification features to the high magnification features, ensuring that prediction performance remains robust. We evaluate MAG-GLTrans across six common computational pathology tasks, including cancer subtyping (TCGA-NSCLC), cancer subtyping in frozen sections (TCGA-NSCLC), prediction of three molecular events (TCGA-Glioma), and lymph node metastasis detection (CAMEYLON16). Our results demonstrate that MAG-GLTrans achieves high-performance classification results (Figure~\ref{fig1}e) on par with state-of-the-art models (CLAM~\cite{lu2021data} and TransMIL~\cite{shao2021transmil}) while reducing computational time by approximately tenfold (Figure~\ref{fig1}g) and cutting communication and storage overhead by twentyfold (Figure~\ref{fig1}h). Furthermore, we highlight the extensibility of MAG (Figure~\ref{fig1}f), showcasing its ability to enhance the efficiency of various CPath architectures and its compatibility with histopathology-specific encoders (e.g., CTransPath~\cite{wang2022transformer}) and foundation models (e.g., UNI~\cite{chen2024towards}) to support low magnification inputs. In addition, the evaluation on a real-world clinical application, computer-assisted telepathology for intraoperative frozen section diagnosis (NSCLC subtyping on TCGA-NSCLC (frozen section)), suggests that MAG-GLTrans is currently the optimal solution for this application due to the low computational and communication costs (Figure~\ref{fig6}). 

\section*{Results}

\begin{figure}[t!]
	\centering
	\includegraphics[width=.9\linewidth]{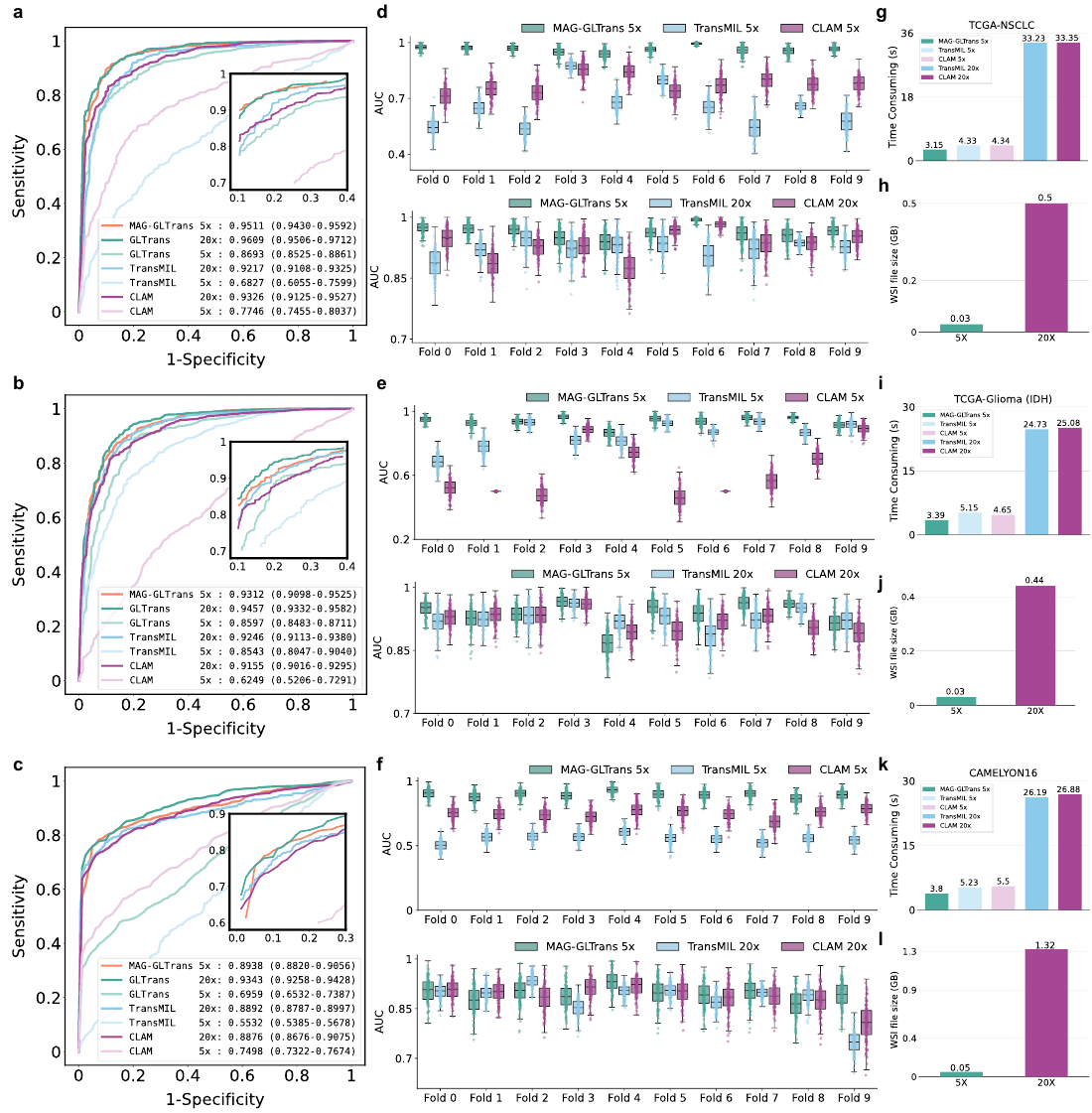}
	\caption{\textbf{Classification performance, computational efficiency and storage burden ($5\times$ versus $20\times$, TCGA-NSCLC, TCGA-Glioma, CAMELYON16).} \textbf{a-f}, the classification performance of the 10-fold cross-validation experiment on three CPath tasks, including subtyping on TCGA-NSCLC (\textbf{a, d}), IDH mutation prediction on TCGA-Glioma (\textbf{b, e}) and lymph node metastasis detection on CAMELYON16 (\textbf{c, f}). \textbf{a-c}, Mean AUC of MAG-GLTrans (5$\times$), GLTrans (5$\times$ and 20$\times$), CLAM (5$\times$ and 20$\times$) and TransMIL (5$\times$ and 20$\times$). \textbf{d-f}, Box plots demonstrate the results of each fold in the 10-fold cross-validation by resampling 500 times. We compare MAG-GLTrans (5$\times$) with two competitors at different magnification levels (5$\times$, top) and (20$\times$, bottom). \textbf{g-l}, Mean computational time (\textbf{g, i, k}) of each model and mean size of the WSI files (\textbf{h, j, l}) on three different datasets.}
	\label{fig2}
\end{figure}

\subsection*{Overall performance of MAG-GLTrans pipeline}
We evaluate the proposed pipeline on three publicly available datasets, including cancer subtyping on TCGA-NSCLC, molecular status predictions on TCGA-Glioma and breast cancer lymph node metastasis detection on CAMELYON16~\cite{bejnordi2017diagnostic}. Specifically, we predicted three important molecular events, TERT mutation, IDH mutation, and 1p19q co-deletion, on the TCGA-Glioma dataset according to the 2021 WHO Classification of Tumors of the Central Nervous System~\cite{10.1093/neuonc/noab106}. For each prediction task, we conduct quantitative comparisons with two outstanding WSI analysis approaches, CLAM~\cite{lu2021data} and TransMIL~\cite{shao2021transmil}, at three different magnification levels ($5\times$, $10\times$ and $20\times$). For each magnification, we first use ImageNet pre-trained ResNet50 as the feature extractor, marked as $20\times$, $10\times$ and $5\times$. For $5\times$ and $10\times$ magnifications, we `upgrade' the extracted features by MAG as $5\times$ (w/ MAG) and $10\times$ (w/ MAG). The patches at different magnification levels share exactly the same image contents but different image resolutions. 10-fold cross-validation is conducted on all the prediction tasks to demonstrate the overall performance. On the TCGA-Glioma and TCGA-NSCLC datasets, all the patients are randomly divided into a training set (80\%), a validation set (10\%), and a test set (10\%). Multiple slides that come from one patient are all sampled into the same set to avoid the data leakage problem. On the CAMELYON16 dataset, since the challenge split the entire dataset into a training set and a test set. The test set remains unchanged, and the original training data is further divided into a training set (80\%) and a validation set (20\%). Besides the classification performance, we also record the runtime of each dataset to demonstrate the computational efficiency. Each dataset is saved at different magnifications to evaluate the storage burden. Additional details regarding the performance of $10\times$ versus $20\times$ are provided in Extended Figure~\ref{supp-fig1}. The performance of TERT mutation and 1p19q codeletion predictions on TCGA-Glioma are provided in Extended Figure~\ref{supp-fig2} and Extended Figure~\ref{supp-fig3}, respectively. The performance of each fold, including AUC, accuracy and F1-score, in different magnification levels is provided in Supplementary materials.


Overall, the classification performance on three datasets is demonstrated in Figure~\ref{fig2}a-c. From the area under the receiver operating characteristic (AUC) results on the TCGA-NSCLC task, we observe that TransMIL and CLAM given $20\times$ magnification inputs, defined as TransMIL ($20\times$) and CLAM ($20\times$), achieved a favorable AUC value of 0.9217 (95\% confidence interval (CI) 0.9108-0.9325) and 0.9314 (95\% CI 0.9125-0.9527), respectively. The computational time per slide is 33.35 seconds (s) for CLAM and 33.23s for TransMIL, while the storage burden per slide is around 0.5 gigabytes (GB). Given $5\times$ inputs, the storage burden and the computational time greatly reduce to around 0.03 GB and 4.3s per slide. Unfortunately, due to the information loss in the low magnification images, the classification performance also drastically decreases to 0.6781 (95\% CI 0.5981-0.7580, -23.90\%) for TransMIL and 0.7746 (95\% CI 0.7455-0.8037, -18.68\%) for CLAM. Thanks to the self-supervised magnification alignment strategy proposed in our study, MAG-GLTrans ($5\times$) is able to bridge the information gap between two magnifications, leading to an exceptional improvement with respect to AUC compared with TransMIL ($5\times$) (0.9511 (95\% CI 0.9430-0.9592) versus 0.6781 (95\% CI 0.5981-0.7580), $P<0.001$) and CLAM ($5\times$) (0.9511 (95\% CI 0.9430-0.9592) versus 0.7746 (95\% CI 0.7455-0.8037), $P<0.001$). The computational time per slide of MAG-GLTrans ($5\times$) is further reduced compared with TransMIL ($5\times$) and CLAM ($5\times$) (3.15s versus 4.33s versus 4.34s). Additionally, MAG-GLTrans ($5\times$) even marginally outperform TransMIL ($20\times$) (0.9511 (95\% CI 0.9430-0.9592) versus 0.9217 (95\% CI 0.9108-0.9325), $P<0.001$) and CLAM ($20\times$) (0.9511 (95\% CI 0.9430-0.9592) versus 0.9314 (95\% CI 0.9125-0.9527), $P<0.001$). 
To evaluate the robustness of the models, we randomly resample the patients in each fold for 500 times. The box-plot of each fold in the 10-fold cross-validation demonstrates that MAG-GLTrans ($5\times$) consistently outperforms TransMIL and CLAM at $5\times$ (Figure~\ref{fig1}d (top)) and $20\times$ (Figure~\ref{fig1}d (bottom)). 

The same observations are found in TCGA-Glioma (IDH) and CAMELYON16. In the TCGA-Glioma (IDH) task, MAG-GLTrans ($5\times$) achieves slightly higher mean AUC compared with TransMIL ($20\times$) (0.9312 (95\% CI 0.9098-0.9525) versus 0.9246 (95\% CI 0.9113-0.9380), $P<0.001$) and CLAM ($20\times$) (0.9312 (95\% CI 0.9098-0.9525) versus 0.9155 (95\% CI 0.9016-0.9295), $P<0.001$), while drastically reduces computational time (3.39s versus 24.73s versus 25.08s) and storage burden from 0.44GB to 0.03GB. In the CAMELYON16 task, MAG-GLTrans ($5\times$) achieves comparable results with TransMIL ($20\times$) (0.8938 (95\% CI 0.8820-0.9056) versus 0.8980 (95\% CI 0.8822-0.9137), $P<0.001$) and CLAM ($20\times$) (0.8938 (95\% CI 0.8820-0.9056) versus 0.8876 (95\% CI 0.8676-0.9075), $P<0.001$), while with much less computational time (3.8s versus 26.19s versus 26.88s), lower storage burden (0.05GB ($5\times$) versus 1.32GB ($20\times$)).

Furthermore, we conduct the same experiment at the $10\times$ magnification, shown in Extended Figure~\ref{supp-fig1}. For example in the CAMELYON16 task, MAG-GLTrans ($10\times$) also achieves the highest mean AUC values compared with TransMIL and CLAM in two magnification levels ($5\times$ and $20\times$). It is not surprising that MAG-GLTrans ($10\times$) slightly outperforms MAG-GLTrans ($5\times$) (0.9108 (95\% CI 0.9048-0.9168) versus 0.8938 (95\% CI 0.8820-0.9056), $P<0.001$), since $10\times$ images lose less information than $5\times$ ones. But the computational time (8.97s versus 3.8s) and storage burden (0.17GB versus 0.05GB) at $10\times$ magnification are higher than the ones at $5\times$ magnification.

In addition, it can be observed that GLTrans proposed in this study substantially outperforms two state-of-the-art (SOTA) WSI analysis architectures at different magnifications in all the five tasks (Figure~\ref{fig2} and Extended Figure~\ref{supp-fig1}). For example in the CAMELYON16 task, GLTrans greatly outperforms TransMIL (0.9343 versus 0.8892, $P<0.001$) and CLAM (0.9343 versus 0.8876, $P<0.001$) at $20\times$. By leveraging the long-range modeling capability of the transformer architecture, GLTrans is ideally suited for WSI analysis with such massive information on the tumor microenvironment. Different from the earlier transformer-based model, like TransMIL, GLTrans introduces global attention and local attention, to preserve the long-range dependency while introducing locality.

\subsection*{MAG can be extended to other WSI analysis architectures}
\begin{figure}[t!]
	\centering
	\includegraphics[width=.9\linewidth]{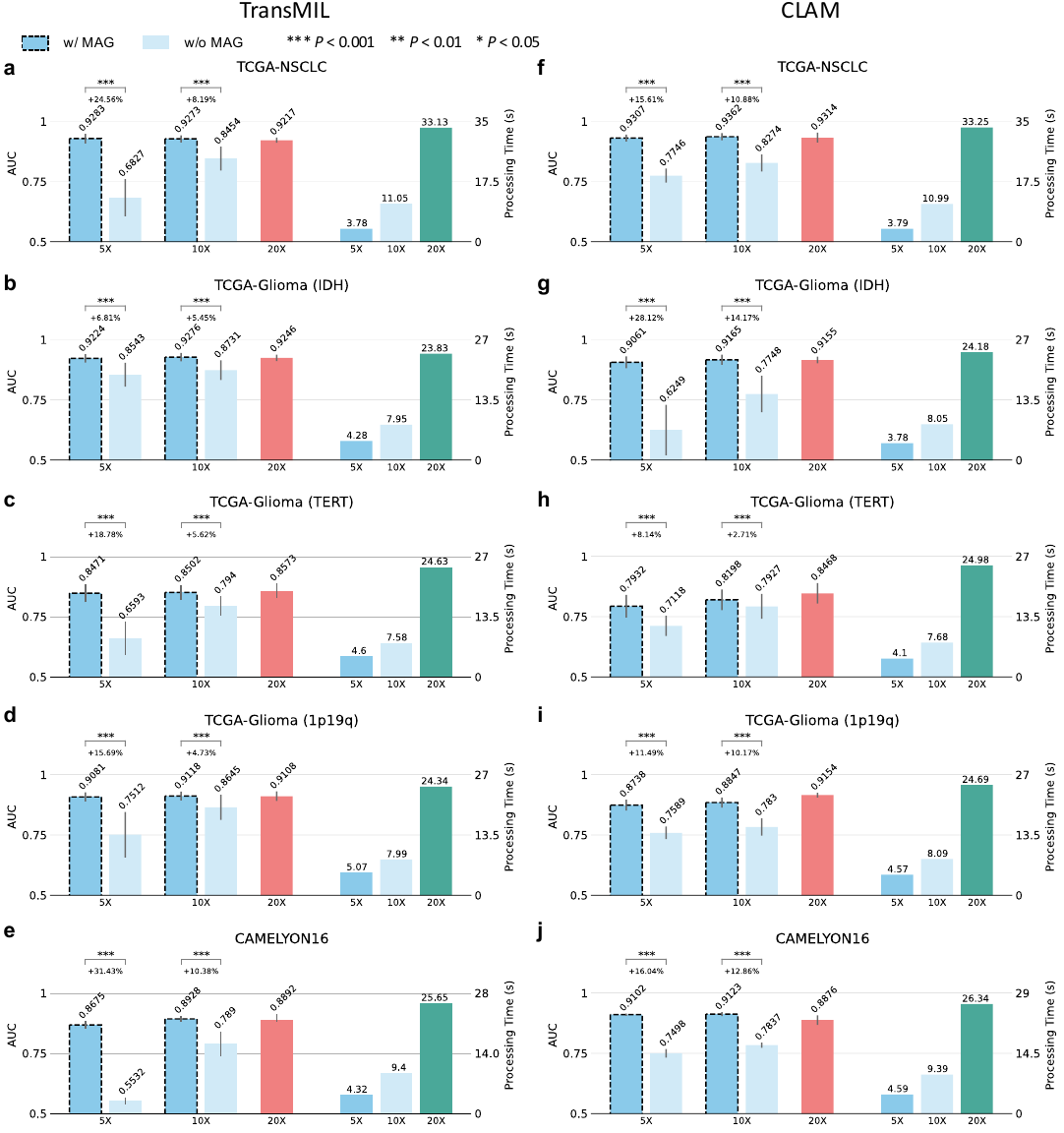}
	\caption{\textbf{Extensibility of MAG to other WSI analysis architectures.} MAG can be applied to any general-purpose WSI analysis architecture to improve computational efficiency. \textbf{a-j}, The plots demonstrate the performance (left) and the computational time (right) of TransMIL (\textbf{a-e}) and CLAM (\textbf{f-j}) at different magnification levels on five CPath tasks. The bars on the left represent the mean AUCs of the 10-fold cross-validation at $5\times$, $10\times$ and $20\times$ with MAG (dash border) and without MAG. The bars on the right show the computational time at different magnification levels. Error bars represent 95\% confidence intervals. The performance increment introduced by MAG is shown on top of the performance bars.}
	\label{fig3}
\end{figure}

Amid the era of deep learning, the proliferation of AI technologies has greatly shortened the life cycles for algorithmic innovations and upgrades. As a result, the extensibility of models has become increasingly critical. In this experiment, we explore the extensibility of our proposed MAG to enhance computational efficiency and communication efficiency for the other WSI analysis architectures, including TransMIL and CLAM. We perform this experiment on five CPath tasks with 10-fold cross-validation, including NSCLC subtyping (TCGA-NSCLC), three molecular status predictions (TCGA-Glioma) and breast cancer lymph node metastasis detection (CAMELYON16). All the experiments are conducted at different input magnification levels ($5\times$, $10\times$ and $20\times$). ImageNet pre-trained ResNet50 is used as the feature extractor for $20\times$ magnification. When training TransMIL and CLAM at low magnification ($5\times$ and $10\times$), the original pipelines remain unchanged except for replacing the feature extractor with our proposed MAG. For each magnification level, a unique MAG model is trained independently.

We observe that the classification performance of both models given low magnification inputs significantly drops in all five CPath tasks when features are extracted using the original pipeline without MAG. For example, the mean AUC values of TransMIL in the NSCLC subtyping task drop from 0.9314 ($20\times$) to 0.8274 ($10\times$) and 0.7746 ($5\times$) shown in Figure~\ref{fig3} (left). When equipped with MAG, the results with low magnification inputs ($5\times$) and ($10\times$) returns to a reasonable level ($0.7746\rightarrow0.9307$ at $5\times$, $P<0.001$, $0.8274\rightarrow0.9362$ at $10\times$, $P<0.001$), which is comparable with the model trained with high magnification inputs ($20\times$). Overall, TransMIL and CLAM equipped with MAG consistently outperform the models without MAG, with the exceptional improvement of mean AUC values of 
(TCGA-NSCLC: $+24.56\%$ [$5\times$], $+8.19\%$ [$10\times$]), 
(TCGA-Glioma (IDH): $+6.81\%$ [$5\times$], $+5.45\%$ [$10\times$]), 
(TCGA-Glioma (TERT): $+18.78\%$ [$5\times$], $+5.62\%$ [$10\times$]), 
(TCGA-Glioma (1p19q): $+15.69\%$ [$5\times$], $+4.73\%$ [$10\times$]) and 
(CAMELYON16: $+31.43\%$ [$5\times$], $+10.38\%$ [$10\times$]) for TransMIL. 
MAG also empowers CLAM with the improvement of mean AUC values of 
(TCGA-NSCLC: $+15.61\%$ [$5\times$], $+10.88\%$ [$10\times$]), 
(TCGA-Glioma (IDH): $+28.12\%$ [$5\times$], $+14.17\%$ [$10\times$]), 
(TCGA-Glioma (TERT): $+8.14\%$ [$5\times$], $+2.71\%$ [$10\times$]), 
(TCGA-Glioma (1p19q): $+11.49\%$ [$5\times$], $+10.17\%$ [$10\times$]) and 
(CAMELYON16: $+16.04\%$ [$5\times$], $+12.86\%$ [$10\times$]). 
This suggests that MAG bridges the information gap and successfully recovers the loss information. Furthermore, we also observe a larger vibration of the performance at $5\times$ magnification (without MAG) than that at $10\times$ magnification (without MAG) due to the different degrees of information loss. Therefore, the increment of the classification performance at $5\times$ magnification that MAG introduces is much larger than the increment at $10\times$ magnification. Another observation is that MAG improves data efficiency when with small-scale data. For example, in CAMELYON16 with only 270 WSIs in the training set, the performance increment for both CLAM and TransMIL is generally larger than the other two datasets with around one thousand WSIs for model training. 
By maintaining a robust classification performance at low magnification levels, MAG brings all the benefits to any WSI analysis pipeline, including computational efficiency, communication efficiency, and a storage-friendly nature. Specifically, it greatly reduces the computational time from around half a minute to only a few seconds, for example from 33.13s to 3.78s for TransMIL in TCGA-NSCLC shown in Figure~\ref{fig3}a (right). Making current and future WSI analysis pipelines feasible and friendly to the clinical scenarios with a demand for fast diagnosis or efficient data transfer, like intraoperative diagnosis or telepathology. 

\begin{figure}[t!]
	\centering
	\includegraphics[width=.8\linewidth]{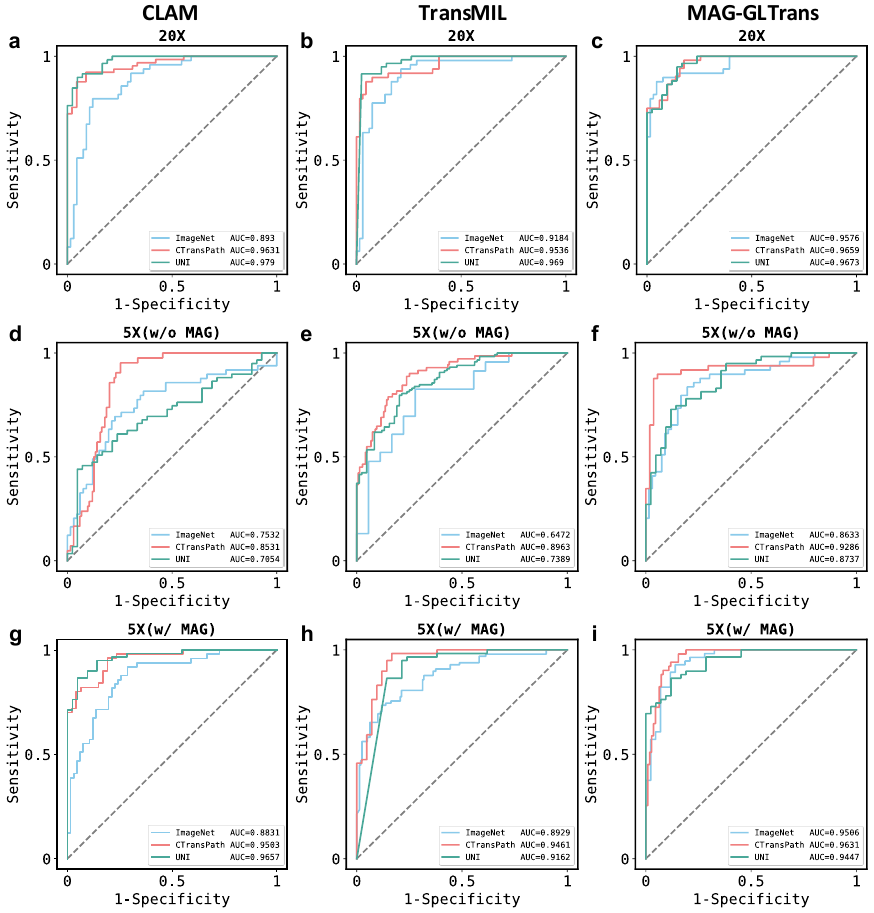}
	\caption{\textbf{Extensibility of MAG to the histopathology-specific encoder.} MAG can be applied to a histopathology-specific encoder or pathology vision foundation model to extract fine-grained domain-specific features from low magnification inputs. Receiver operating characteristic (ROC) curves demonstrate the performance of CLAM (\textbf{a, d, g}), TransMIL (\textbf{b, e, h}) and MAG-GLTrans (\textbf{c, f, i}) on TCGA-NSCLC using MAG trained with domain-unspecific (ImageNet pre-trained ResNet50) and domain-specific models (TCGA pre-trained CTransPath), respectively. From top to bottom: ROC curves at $20\times$, $5\times$ without MAG and $ 5\times$ with MAG.}
	\label{fig4}
\end{figure}

\subsection*{MAG can be extended to the histopathology-specific encoder}
Histopathology-specific encoders or foundation models~\cite {wang2022transformer,chen2024towards} have been proven to be much more effective for WSI analysis in various CPath tasks compared with the feature extractor pre-trained on natural images. MAG, as a self-supervised model that can align the learned knowledge and restore the information loss between two magnifications, is capable of being extended to the histopathology-specific encoder, making it possible for low magnification inputs. Here, we investigate the extensibility of MAG on the earliest histopathology-specific model, CTransPath~\cite{wang2022transformer} as well as the recently proposed pathology vision foundation model, UNI~\cite{chen2024towards}. We conduct this experiment on the NSCLC subtyping task (TCGA-NSCLC). CLAM, TransMIL and MAG-GLTrans are trained at the high magnification $20\times$ and the low magnification $5\times$ (with and without MAG). 

As we expected, the histopathology-specific pre-trained feature extractors at $20\times$ brings appropriate improvements for all three models (Figure~\ref{fig4}a-c) compared with the ImageNet pre-trained feature extractor with respect to AUC (CTransPath: CLAM=0.9631, $+7.01\%$, TransMIL=0.9536, $+3.52\%$, MAG-GLTrans=0.9659, $+0.83\%$) and (UNI: CLAM=0.9790, $+8.06\%$, TransMIL=0.9690, $+5.06\%$, MAG-GLTrans=0.9673, $+0.97\%$). When reducing the magnification level to $5\times$ without the support of MAG, the classification performance (Figure~\ref{fig4}d-f) with three different encoders of the three WSI analysis models decreases due to the information loss. Even though CTransPath without MAG also faces an obvious performance degradation at $5\times$, applying a histopathology-specific feature extractor is still a better solution compared with the ImageNet pre-trained one. When training MAG on three different encoders at the $5\times$ magnification level, the performance of three WSI analysis architectures returns to a favorable level, even outperform the model trained at $20\times$ with the ImageNet pre-trained feature extractor. The model-agnostic nature of MAG is a highly significant feature that enables it to be seamlessly integrated with any top-tier, histopathology-specific pre-trained model, resulting in minimal information loss. This capability substantially enhances computational efficiency and communication efficiency, all without compromising the model's classification performance.

In addition, by summarizing the results in Figure~\ref{fig4} horizontally, we observe that the prediction model (GLTrans) introduced in this study generally outperforms CLAM and TransMIL in any magnification and experimental configuration, which demonstrates that GLTrans is a superior WSI analysis architecture with respect to the capability of spatial and local information modeling.

\begin{figure}[!t]
	\centering
	\includegraphics[width=.985\linewidth]{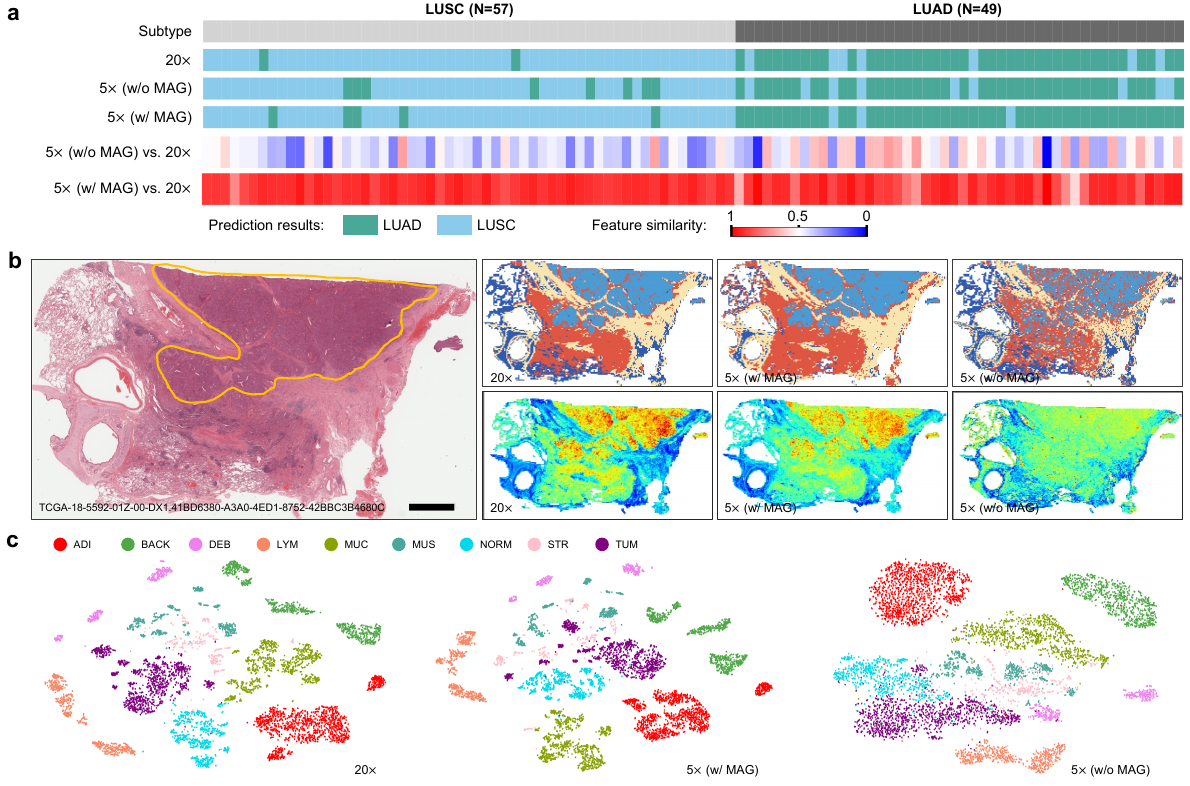}
	\caption{\textbf{Visualization of the features extracted by MAG (TCGA-NSCLC).} \textbf{a}, The prediction results (top) of one fold on the subtyping task (TCGA-NSCLC, $n=106$) under different magnification levels. Patients are sorted by the ground truth labels. Heatmaps (bottom) demonstrate the similarity of the features between high magnification level and low magnification level (with and without MAG). We show the feature similarity in slide level, which is the mean of all the patches. Red represents higher feature similarity. \textbf{b}, A sample case WSI (left) of LUSC with pathologist-annotated carcinoma region (yellow border), visualization of the unsupervised clustering of the patch-level features (top right) and the heatmap (bottom right) at $20\times$, $5\times$ with MAG and $5\times$ without MAG. \textbf{c}, T-SNE visualization of the image embeddings from three different models in the Kather colon dataset~\cite{kather} with nine tissue types.}
	\label{fig5}
\end{figure}

\subsection*{MAG features are reliable and interpretable}
Model interpretability and reliability are the most important natures when applying AI models to real-world clinical scenarios. As a feature extractor given low magnification inputs, MAG faces a significant challenge of information loss, potentially undermining the model's reliability and interpretability. In this experiment, we investigate the feature quality with and without MAG, by visualizing them at cohort, slide and patch levels. The visualization of NSCLC subtyping (TCGA-NSCLC) and breast cancer lymph-node metastasis detection (CAMELYON16) are shown in Figure~\ref{fig5} and Extended Figure~\ref{supp-fig4} respectively.

To evaluate and visualize whether MAG is able to restore the information loss at the low magnification, we quantify the similarity of the features between low magnification $5\times$ (with and without MAG) and high magnification $20\times$. Specifically, we calculate a slide-level similarity score by averaging the similarity scores (normalized Euclidean distances) across all patches. The heatmaps depicted in Figure~\ref{fig5}a indicate that the features extracted from the low magnification images using MAG are significantly more similar to those from the high magnification images than the features extracted at low magnification without MAG. This finding suggests that MAG can refine the features extracted from low magnification images, effectively bridging a substantial information gap of up to 16 times. With those more reliable features, the rate of misclassify has noticeably decreased at $5\times$ magnification using MAG compared with the ones in the $5\times$ without MAG. We also observe that the misclassified samples at $5\times$ without MAG tend to exhibit lower feature similarities, typically falling below the threshold of 0.5. For the CAMELYON16 dataset with fewer training samples and larger color variations, unreliable features can severely impair the prediction model, resulting in significantly degraded classification performance with more misclassified samples (Extended Figure~\ref{supp-fig4}).

To further visualize the robustness and the representativeness of MAG features, and to evaluate whether those features deliver enough semantic information, we present a slide-level visualization in Figure~\ref{fig5}b. We first apply an unsupervised clustering approach (K-means) on the patch-level features and group them into four clusters. The cluster maps shown in the top-right row demonstrate that the features extracted from high magnification patches are capable of capturing and reflecting the morphological characteristics accurately with a clear boundary between tumor and normal tissues and a high agreement with the boundary drawn by the pathologist. When reducing the magnification to $5\times$, the tissue boundaries become unclear and the clustering map is much noisier than the high magnification one. This observation indicates that reducing image resolution can greatly harm the semantic information, resulting in unstable prediction results. By bridging information gap using MAG, the clustering map of $5\times$ with MAG maintains a high visual consistency with the one of $20\times$. Next, we plot the heatmaps of the WSI analysis model GLTrans in the top-right row in Figure~\ref{fig5}b. The color intensity of the heatmaps correlates with the confidence of the classification outcome for that particular patch. When without MAG, the lower magnification features lead to less effective heatmaps, indicating that the feature representation of the predictive models is impaired by these unstable and less detailed features. Applying MAG significantly enhances the effectiveness of features, resulting in heatmaps that are much more consistent with those at higher magnification. The feature space of the image embeddings (Figure~\ref{fig5}c) also demonstrates the effectiveness of MAG.

\begin{figure}[ht!]
	\centering
	\includegraphics[width=\linewidth]{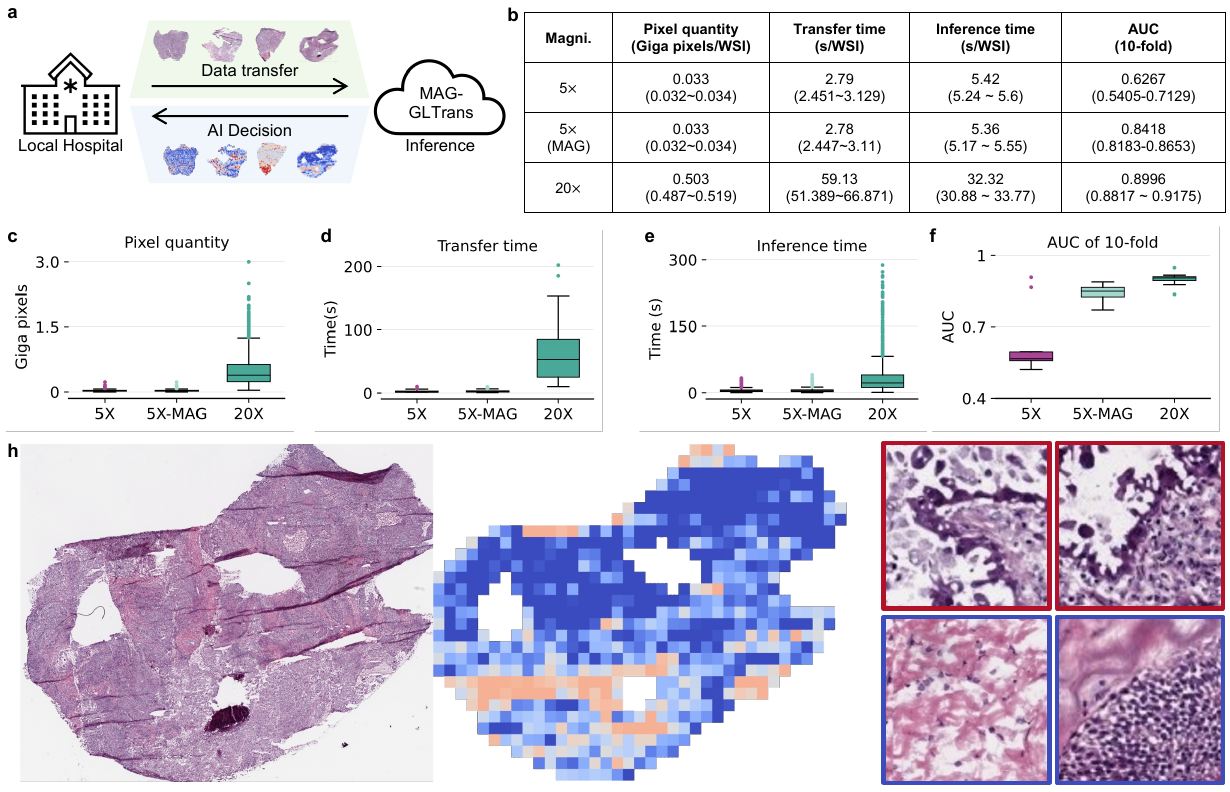}
	\caption{\textbf{MAG-GLTrans on telepathology for intraoperative frozen section diagnosis.} \textbf{a}, Local hospital transfers low magnification WSIs to the cloud server and obtains a diagnostic result and an interpretable heatmap in a only few seconds. \textbf{b-f}, Statistical results (\textbf{b}) and the box plots at different magnification levels with respect to the pixel quantity (\textbf{c}), data transfer time (\textbf{d}), model inference time (\textbf{e}) and the diagnostic performance of AUC (\textbf{f}). \textbf{h}, A sample case and its corresponding heatmap as well as the patches with high confidence (red border) and low confidence (blue border).}
	\label{fig6}
\end{figure}

\subsection*{Communication-efficient Telepathology for intraoperative frozen section diagnoses}
We evaluate the potential of MAG-GLTrans for computer-assisted telepathology in intraoperative frozen section diagnosis, a critical need for hospitals lacking on-site expert pathologists, with a focus on classification performance, computational efficiency, and communication efficiency. This investigation is conducted on the NSCLC subtyping task using the TCGA-NSCLC dataset (frozen sections), which includes 1067 frozen section WSIs of LUAD and 1100 WSIs of LUSC. A 10-fold cross-validation approach is employed, with all patients randomly partitioned into a training set (80\%), a validation set (10\%), and a test set (10\%). To prevent data leakage, all slides from a single patient are assigned to the same set. Classification performance is assessed using the AUC metric. For each WSI, we store each magnification level as a separate file to accurately measure storage consumption, data transfer time, and model inference time across different magnifications. Transmission performance is evaluated under a 200 Mbps bandwidth connection during typical working hours (8:00-18:00) to simulate real-world clinical conditions.

For local hospitals lacking on-site expert pathologists, particularly in underdeveloped regions or countries, we propose an efficient computer-assisted telepathology (CATP) framework tailored for frozen section diagnosis (Figure~\ref{fig6}a). This framework enables local hospitals to transmit low-magnification whole slide images (WSIs) to cloud servers and receive computer-assisted diagnostic results alongside interpretable heatmaps within seconds. To rigorously assess the framework's efficacy, we conducted an ablation study, as depicted in Figure~\ref{fig6}b. The results reveal that, as anticipated, higher magnification WSIs require significantly more storage due to their substantially larger pixel counts (20$\times$: 0.503 Gigapixels/WSI vs. 5$\times$: 0.033 Gigapixels/WSI), leading to increased data transfer and model inference times. On average, processing a single WSI takes approximately one minute (59.13 seconds) for file transmission and half a minute (32.32 seconds) for model inference. In real-world clinical settings, these times can vary considerably due to factors such as network bandwidth, latency fluctuations, computational resource availability, concurrent user load, and other environmental variables.
Reducing magnification yields remarkable efficiency gains, improving processing speed by more than tenfold. Specifically, data transfer time drops dramatically from 59.13 seconds per WSI at 20$\times$magnification to just 2.79 seconds at 5$\times$, while model inference time decreases from 32.32 seconds to around 5 seconds per WSI. However, this optimization comes at a significant cost to diagnostic accuracy, as indicated by the sharp decline in AUC values from 0.8995 at 20$\times$ to 0.6267 at 5$\times$, severely limiting clinical utility. Our proposed MAG-GLTrans framework effectively addresses this trade-off by maintaining high predictive performance (AUC: 0.8418) with only marginal degradation compared to 20$\times$ magnification, while simultaneously mitigating the drawbacks of magnification reduction. By significantly reducing both data transfer and model inference times, this innovative CATP framework greatly enhances the clinical feasibility of intraoperative frozen section diagnosis.

\section*{Discussion}
We introduce MAG-GLTrans, a general-purpose WSI analysis architecture, designed for a wide range of computational pathology applications, including cancer subtyping, metastasis detection, and mutation prediction. Beyond its exceptional classification capabilities, MAG-GLTrans stands out for its computational efficiency, storage-friendly design, extensibility, and communication efficiency. These attributes, whose importance is first emphasized in this study, are often overlooked in the field of computational pathology (CPath). The primary obstacle to advancing CPath towards greater efficiency lies in the fact that current WSI analysis architectures typically process WSIs at high magnifications~\cite{hosseini2024computational}. This approach is constrained by the limitations of existing computer hardware. While directly reducing image resolution can improve efficiency, it often comes at the cost of performance due to information loss. MAG addresses this challenge by bridging the information gap and extracting fine-grained features from low-magnification images, thereby achieving an optimal balance between performance and efficiency. Through 10-fold cross-validation experiments on five CPath tasks, our model demonstrates robust classification performance at 5$\times$magnification, comparable to two state-of-the-art WSI analysis architectures operating at 20$\times$, while reducing computational time by up to 10.7 times, storage overhead by up to 28.6 times, and communication time by up to 21.5 times. We believe this represents a significant and milestone finding that brings CPath technology one step closer to real-world clinical adoption. For instance, a recent study highlights that reviewing WSIs results in a median overall 19\% decrease in efficiency per case compared to glass slides~\cite{hanna2019whole}. This inefficiency is further exacerbated when pathologists must wait tens of seconds for a decision on a single slide from CPath systems. Such delays are particularly pronounced when analyzing multiple hematoxylin and eosin (H\&E) stained diagnostic slides and immunohistochemistry slides, a common scenario for patients with larger tumors. Even if WSIs are processed offline before pathologist review, hospitals with high patient volumes would require substantial computational resources. MAG-GLTrans addresses this issue by reducing computation turnaround time to just a few seconds, making the computational process virtually seamless for pathologists.

In addition to enhancing local pathology diagnosis, we also investigate the potential of MAG-GLTrans in enabling telepathology (TP), a field of pathology that utilizes telecommunication technologies to facilitate the transfer of digital pathology slides across geographically distant locations~\cite{weinstein2001telepathology}. Intraoperative frozen section diagnosis stands out as one of the most promising applications of TP, particularly in settings where pathologists are scarce, such as hospitals without on-site pathology expertise~\cite{wellnitz2000reliability, evans2009primary, 10.5858/arpa.2022-0261-OA}. Integrating computational pathology techniques to enable computer-assisted telepathology (CATP) for intraoperative diagnosis has the potential to improve diagnostic accuracy~\cite{nam2020introduction}. However, the unpredictable data transfer times under complex network conditions and the prolonged inference times associated with processing gigapixel WSIs pose significant challenges, potentially increasing turnaround times and compromising diagnostic efficiency~\cite{hollon2020near}. These issues highlight the need for further validation of CATP's feasibility.
Our evaluation on the frozen section diagnosis dataset (TCGA-NSCLC (frozen sections)) demonstrates that MAG-GLTrans can significantly reduce the turnaround time required for both data transfer and computation.

Due to the absence of domain-specific encoders before the emergence of foundation models, earlier WSI analysis pipelines typically relied on ImageNet pre-trained models to extract patch-level features as a compromise solution. However, with the accumulation of large-scale histopathology datasets and advancements in computational resources, self-supervised learning on histopathology datasets~\cite{wang2022transformer} and the development of visual foundation models~\cite{huang2023visual,chen2024towards,lu2024visual} have led to significant progress in computational pathology. Notably, histopathology-specific encoders such as CTransPath~\cite{wang2022transformer}, UNI~\cite{chen2024towards}, PLIP~\cite{huang2023visual}, Prov-GigaPath~\cite{xu2024whole}, MUSK~\cite{xiang2025vision} and Virchow~\cite{vorontsov2024foundation} have demonstrated superior effectiveness across various CPath tasks compared to encoders trained on natural images.
To preserve fine-grained information, most existing histopathology-specific encoders are still trained on high-magnification images. MAG, however, can be extensible to any histopathology-specific encoder in a self-supervised manner, enabling the extraction of fine-grained features even from low-magnification inputs. Experimental results indicate that MAG, when trained with a histopathology-specific encoder, further enhances classification performance compared to MAG trained with ImageNet. Additionally, due to its plug-and-play nature, MAG can be seamlessly integrated into other WSI analysis architectures, inheriting their strengths while improving their efficiency. For example, when MAG is applied as the feature extractor for CLAM, the performance  remain largely unaffected even at low magnification. This exceptional extensibility of MAG significantly extends its lifecycle, particularly in light of the rapid pace of algorithmic advancements and iterations in the field.

While the MAG-GLTrans model has demonstrated remarkable efficiency and accuracy in handling low-magnification inputs, several areas warrant further exploration to unlock its full potential. Firstly, we hypothesize that the model's capabilities could extend beyond its current resolution limits. By applying the model to even lower magnifications, such as 2.5$\times$, it may be possible to achieve further improvements in computational speed and communication efficiency. However, this approach requires a careful balance between reducing image resolution and preserving essential diagnostic details. Whether this trade-off will compromise the model's effectiveness at lower magnifications remains an open question that merits thorough investigation. Additionally, due to the lack of large-scale datasets, MAG was trained on a limited number of WSIs as a compromise. Investigating its performance on a larger-scale dataset would be highly valuable, as it could further enhance the model's generalization capabilities and robustness. This exploration could provide deeper insights into the scalability and effectiveness of MAG when applied to more diverse and extensive histopathology data, potentially unlocking even greater performance improvements and broader applicability in computational pathology.

In summary, to the best of our knowledge, MAG-GLTrans represents the first approach in computational pathology that simultaneously addresses classification capability, computational efficiency, storage requirements, and communication burden. By significantly enhancing efficiency, MAG-GLTrans renders the computational process virtually imperceptible to pathologists, marking a critical step toward integrating computational pathology into real-world clinical practice. This advancement is particularly impactful in applications such as computer-assisted telepathology (CATP) for intraoperative frozen section diagnosis, where processing speed, accuracy, and resource efficiency are paramount.

\section*{Methods}
\subsection*{MAG-GLTrans}
MAG-GLTrans is a computation-efficient, communication-efficient and storage-friendly WSI analysis pipeline. The same with existing mainstream WSI analysis pipelines, it consists of two main modules, a feature extractor (MAG) and a predictor (GLTrans). The major difference is that MAG uses low magnification inputs and the conventional encoders use high magnification ones.\\ 

\noindent \textit{Magnification Alignment (MAG)}\\
MAG is a self-supervised training strategy that enables the expansion of an encoder, initially trained on high magnification images, to effectively handle low magnification inputs. The main concept of MAG is to minimize the information gap between two magnification levels by aligning their respective feature spaces. To achieve this, two models with identical network structures are employed, including a well-trained feature extractor for high magnification level and a feature extractor for low magnification level (Figure~\ref{fig1}b). $L1$ loss is applied to make the features extracted from different encoders as similar as possible, from the shallow layers to the deep layers. 
\begin{equation} 
	\label{MAG_loss}
	\mathcal{L}_{MAG} = \sum_{i=1}^{n} {|X_{i}^{H} - X_{i}^{L}|}x
\end{equation} 
where $X_{i}^{H}$ and $X_{i}^{L}$ denote the feature maps of the $i$-th block extracted from the high magnification encoder and the low magnification encoder, respectively. $n$ is the number of blocks. For example in ResNet50, $n$ is 5, where 1 to 4 correspond to layer1 to layer4 and 5 represents the output of the global average pooling layer. During MAG training, we only update the low magnification encoder and freeze the high magnification one. The image resolution of the patches at different magnifications is 256$\times$256 ($20\times$), 128$\times$128 ($10\times$) and 64$\times$64 ($5\times$). The patches at the same location under different magnifications share exactly the same image content. For the WSIs without predefined specific magnification levels, we directly downsample the patches to the corresponding magnification level. Theoretically, using as large a dataset as possible to train a model can yield a diverse and robust feature extractor. Nonetheless, due to the constraints on computational resources, we train a distinct model for each dataset. 3,293,000, 12,244,980 and 2,821,910 non-overlapping histopathological image patches are extracted from TCGA-NSCLC, TCGA-Glioma and CAMELYON16 are used to train MAG, respectively. The model parameters are updated using the Adam optimizer with an initial learning rate of 0.0001. The learning strategy is linear decay with a decay coefficient of 0.9. Since MAG is trained in a self-supervised manner, the training procedure is performed without any manual intervention or labeling burden. In addition, the model-agnostic nature allows MAG to be applied to any well-trained encoder, for example ImageNet pre-trained ResNet, histopathology-specific encoder CTransPath or other pathology foundation models.\\

\noindent \textit{Global-local Transformer (GLTrans)}\\
Due to the complex tumor microenvironment, formulating spatial information for long-range dependencies is crucial for WSI analysis. Therefore, we introduce a novel WSI analysis architecture named the Global-Local Transformer (GLTrans) (Figure~\ref{fig1}a). GLTrans is a transformer-based approach that takes patch embeddings as input. Each patch embedding is regarded as a patch token and is formulated in a multiple instance learning manner. Unlike conventional transformer and multiple instance learning (MIL) based models, such as TransMIL~\cite{shao2021transmil}, GLTrans introduces local and global attention mechanisms in parallel. The global attention mechanism, using multi-head self-attention, constructs the spatial relationships among all patch instances, while the local attention mechanism, applied through instance-level dilated convolution, introduces local contextual information. The dilated rate in local attention is $\{1, 3, 5\}$.
GLTrans uses a cross-entropy loss function for binary classification tasks, defined by equation~\ref{GLT Loss}.
\begin{equation}
	\label{GLT Loss}
	\mathcal{L}_{GLTrans} = -\sum_{i=1}^{n}{y \, log \,\hat{y}}
\end{equation}
where $\hat{y}$ is the output probability, $y$ is the ground truth label.
We randomly initialize the model parameters. The optimizer is Adam and uses the LookAhead mechanism with an initial learning rate of $1 \times 10^{-4}$ and a weight decay of $1 \times 10^{-5}$.
The F1 scores of the validation set are monitored during training. During training, the instances in the input package are randomly disordered to serve as a form of data augmentation, while they are arranged in the order of the feature extraction phase during model inference.

\subsection*{Public datasets}

\noindent \textit{TCGA-NSCLC diagnostic and frozen sectioned data.}\\
We used H\&E stained diagnostic WSIs sourced from the repository of TCGA-NSCLC (956 cases with 1,053 slides) which includes two projects TCGA-LUAD (478 cases with 541 slides) and TCGA-LUSC (478 cases with 512 slides). Since not all the WSIs have the same magnification levels, we applied image downscaling to generate the missing magnification level from the high magnification one. The TCGA-NSCLC cohort was split into 80:10:10 as the training, validation and test sets for 10-fold cross validation. This dataset was formulated as a binary classification task to predict the subtypes (LUAD versus LUSC).

The frozen sectioned slides sourced from the repository of TCGA-NSCLC were used for the experiment of telepathology for intraoperative frozen section diagnosis, which included 1,009 cases with 2,167 slides. Specifically, the TCGA-LUAD consists of 514 cases with 1,067 slides (244 normal and 823 tumor slides). the TCGA-LUSC consists of 495 cases with 1,100 slides (347 normal and 753 tumor slides). We construct two binary classification tasks, one for the malignancy prediction (normal versus tumor) and another for the subtype prediction (LUAD versus LUSC). The malignancy prediction task consists of 591 frozen sectioned slides with only normal tissue and 1576 frozen sectioned slides with malignant tumors. The subtype prediction task includes 753 LUSC slides and 823 LUAD slides.\\

\noindent \textit{TCGA-Glioma.}\\
The WSIs sourced from the repository of TCGA-Glioma (880 cases with 1,704 slides) consists of two projects, TCGA-GBM (389 cases with 860 slides) and TCGA-LGG (491 cases with 844 slides). In this study, we predict three important molecular alteration events (IDH mutation, 1p/19q codeletion and TERT promoter mutation). The prediction of each event is formulated as a binary classification task with the labels extracted from the supplemental file of the previous study~\cite{ceccarelli2016molecular}. Due to the absence of comprehensive molecular profiling data in some cases, we have excluded both the cases and the slides that lack specific molecular status information for each molecular prediction task. The IDH prediction task consists of 802 cases (IDH wild-type: 381, IDH mutant: 421) with 1,539 slides (IDH wild-type: 758, IDH mutant: 781). The 1p/19q codeletion prediction task consists of 866 cases (1p/19q codeletion: 704, 1p/19q non-codeletion: 162) with 1,684 slides (1p/19q codeletion: 327, 1p/19q non-codeletion: 1,357). The TERT promoter prediction task consists of 301 cases (TERT promoter wild-type: 151, TERT promoter mutant: 150) with 574 slides (TERT promoter wild-type: 265, TERT promoter mutant: 309). The datasets of all three prediction tasks were split into 80:10:10 as the training, validation and test sets for 10-fold cross validation, respectively.\\

\noindent \textit{CAMELYON16.}\\
CAMELYON16 is a publicly available dataset for breast cancer lymph node metastasis detection which consists of 270 slides (160 normal and 110 metastasis) in the training set and 129 slides (80 normal and 49 metastasis) in the test set. In the 10-fold cross-validation experiment, we split the original training set into a training set (80\%) and a validation set (20\%). And the model was evaluated on the held-out test set.

\subsection*{Computational hardware and software}
We used Python (v3.10.0) and Pytorch (v2.1.2) to build all deep learning models, experiments and analyses in this study. The ImageNet pre-trained weights of ResNet50 (\url{https://download.pytorch.org/models/resnet50-0676ba61.pth}) and histopathology images pre-trained weights of CTransPath~\cite{wang2022}  (\url{https://github.com/Xiyue-Wang/TransPath}) and UNI~\cite{chen2024towards} (\url{https://github.com/mahmoodlab/UNI}) were used to train MAG. The source codes of CLAM (\url{https://github.com/mahmoodlab/CLAM}) and TransMIL (\url{https://github.com/szc19990412/TransMIL}) were downloaded from the corresponding GitHub repositories. All the models, including MAG-GLTrans, CLAM and TransMIL, as well as the downstream experiments were trained and conducted on a server equipped with a single NVIDIA GeForce RTX 4090 GPU with 256 GB of running memory. Reading of WSIs was done using the openslide-python (v1.3.1) and pyvips (v2.2.1), and writing of the images was done using Opencv-python (v4.8.1). We used the k-mean algorithm in Scikit-learn (v1.3.2) to perform patch-level clustering. The box plots, radar plots, heatmaps, histograms, scatter plots, ROC curves and the patch attention heatmap were drawn by the Python matplotlib (v3.8.2).

\subsection*{Data availability}
The data used in this study are all from the open-source platform. All the TCGA whole-slide data and labels can be accessed through the NIH genomic data commons \url{(https://portal.gdc.cancer.gov)}. The WSIs of TCGA-NSCLC and TCGA-NSCLC (frozen section) were downloaded from two projects of TCGA-LUAD \url{(https://portal.gdc.cancer.gov/projects/TCGA-LUAD)} and TCGA-LUSC \url{(https://portal.gdc.cancer.gov/projects/TCGA-LUSC)}. The WSIs of TCGA-Glioma were downloaded from two projects of TCGA-GBM \url{(https://portal.gdc.cancer.gov/projects/TCGA-GBM)} and TCGA-LGG \url{(https://portal.gdc.cancer.gov/projects/TCGA-LGG)}. The data of CAMELYON16 is publicly available via the Grand Challenge platform at \url{https://camelyon16.grand-challenge.org}.

\subsection*{Code availability}
MAG-GLTrans can be accessed at \url{https://github.com/Bingchao-Zhao/MAG-GLTrans}, including the model weights and the relevant source code. 

\appendix
\setcounter{figure}{0}
\renewcommand{\thefigure}{\thesection\arabic{figure}}
\renewcommand{\figurename}{Extended Figure}

\begin{figure}[htp]
	\centering
	\includegraphics[width=\linewidth]{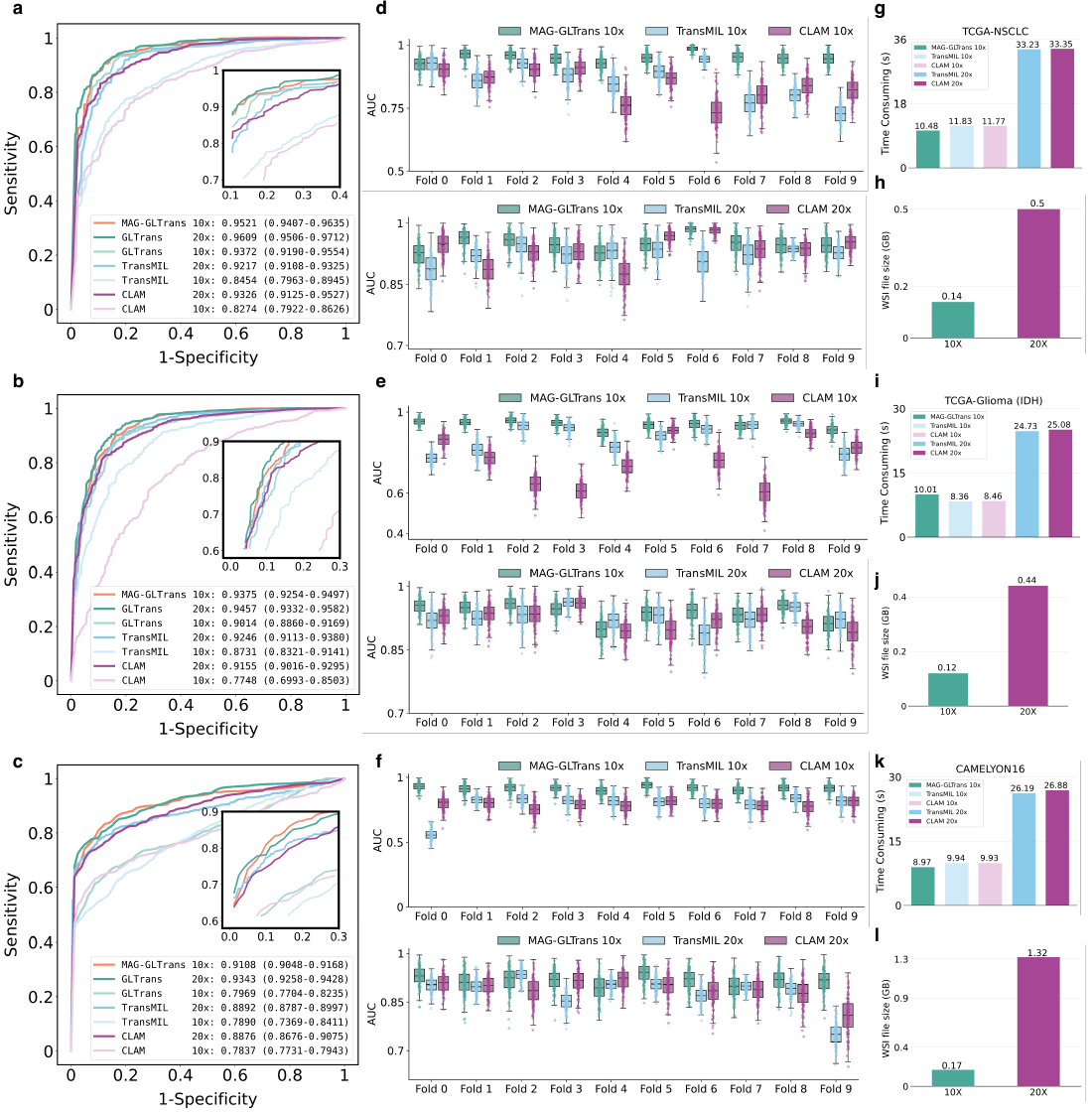}
	\caption{\textbf{Classification performance, computational efficiency and storage burden ($10\times$ versus $20\times$, TCGA-NSCLC, TCGA-Glioma, CAMELYON16).} \textbf{a-f}, the Classification performance of the 10-fold cross-validation experiment on three CPath tasks, including subtyping on TCGA-NSCLC (\textbf{a, d}), IDH mutation prediction on TCGA-Glioma (\textbf{b, e}) and lymph node metastasis detection on CAMELYON16 (\textbf{c, f}). \textbf{a-c}, Mean AUC of MAG-GLTrans ($10\times$), GLTrans ($10\times$ and $20\times$), CLAM ($10\times$ and $20\times$) and TransMIL ($10\times$ and $20\times$). \textbf{d-f}, Box plots demonstrate the results of each fold in the 10-fold cross-validation by resampling 500 times. We compare MAG-GLTrans ($10\times$) with two competitors at different magnification levels ($10\times$, top) and ($20\times$, bottom). \textbf{g-l}, Mean computational time (\textbf{g, i, k}) of each model and mean size of the WSI files (\textbf{h, j, l}) on three different datasets.}
	\label{supp-fig1}
\end{figure}

\begin{figure}[!]
	\centering
	\includegraphics[width=\linewidth]{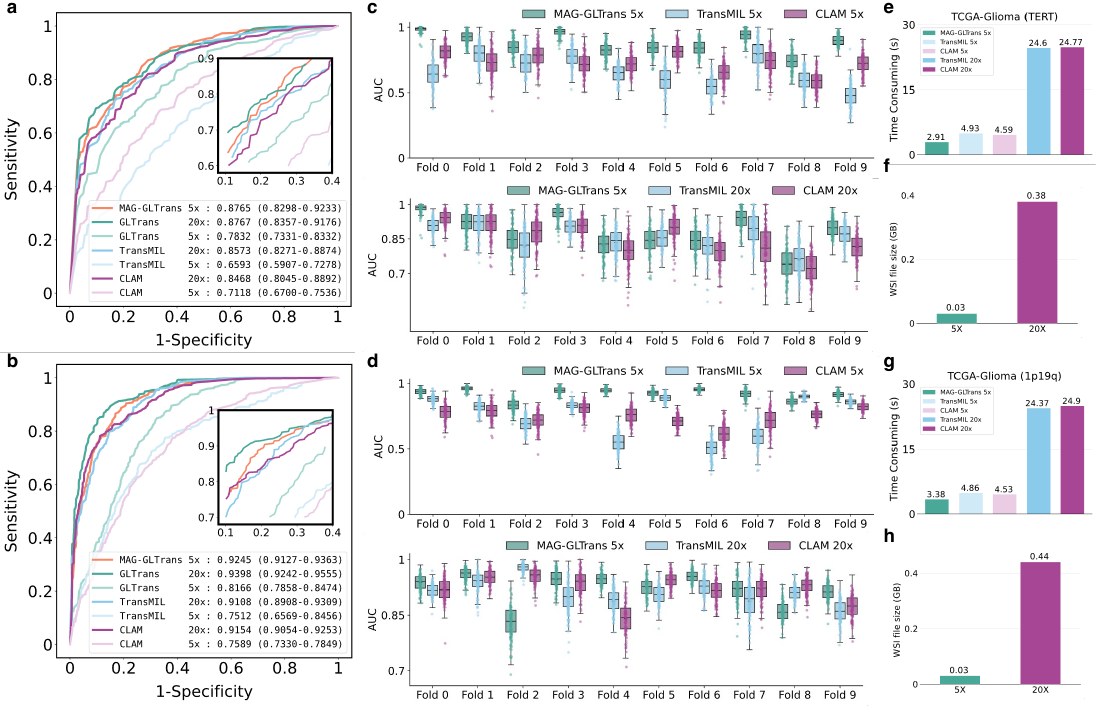}
	\caption{\textbf{Classification performance, computational efficiency and storage burden ($5\times$ versus $20\times$, TCGA-Glioma).} \textbf{a-d}, Classification performance of the 10-fold cross-validation experiment on TERT mutation prediction (\textbf{a, c}) and 1p19q codeletion prediction (\textbf{b, e}) on TCGA-Glioma. \textbf{a-b}, Mean AUC of MAG-GLTrans (5$\times$), GLTrans (5$\times$ and 20$\times$), CLAM (5$\times$ and 20$\times$) and TransMIL (5$\times$ and 20$\times$). \textbf{c-d}, Box plots demonstrate the results of each fold in the 10-fold cross-validation by resampling 500 times. We compare MAG-GLTrans (5$\times$) with two competitors at different magnification levels (5$\times$, top) and (20$\times$, bottom). \textbf{e-h}, Mean computational time (\textbf{e, g}) of each model and mean size of the WSI files (\textbf{f, h}) in two different tasks.}
	\label{supp-fig2}
\end{figure}

\begin{figure}[!]
	\centering
	\includegraphics[width=\linewidth]{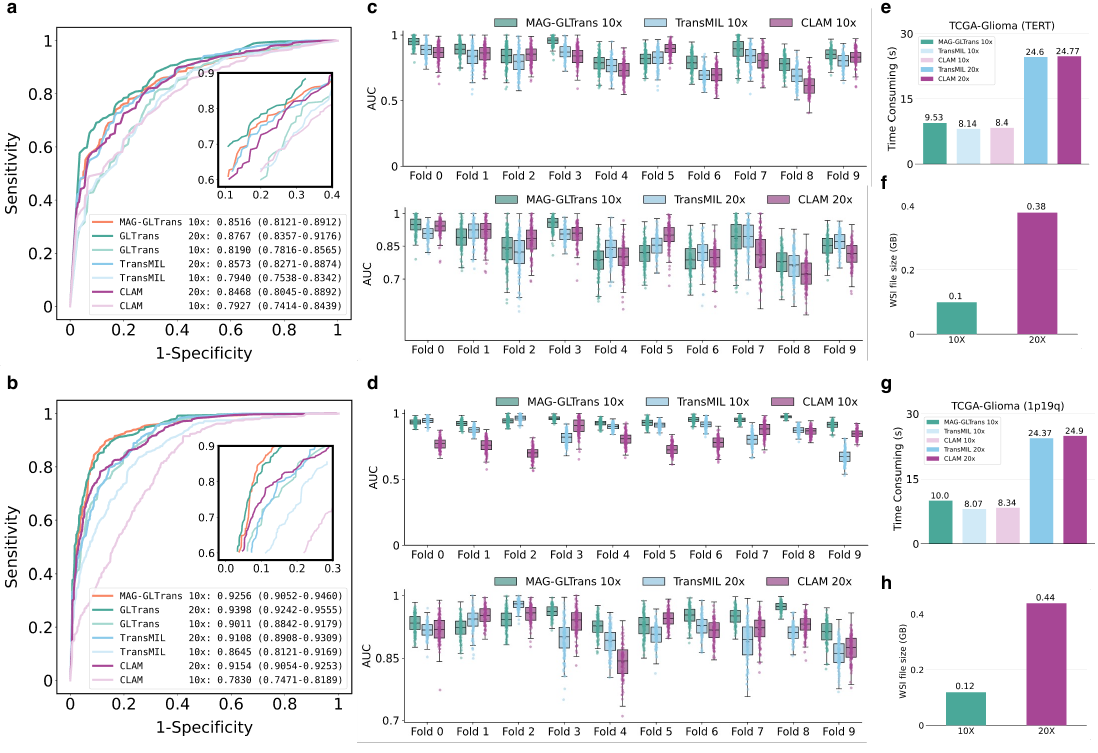}
	\caption{\textbf{Classification performance, computational efficiency and storage burden ($10\times$ versus $20\times$, TCGA-Glioma).} \textbf{a-d}, Classification performance of the 10-fold cross-validation experiment on TERT mutation prediction (\textbf{a, c}) and 1p19q codeletion prediction (\textbf{b, e}) on TCGA-Glioma. \textbf{a-b}, Mean AUC of MAG-GLTrans ($10\times$), GLTrans ($10\times$ and $20\times$), CLAM ($10\times$ and $20\times$) and TransMIL ($10\times$ and $20\times$). \textbf{c-d}, Box plots demonstrate the results of each fold in the 10-fold cross-validation by resampling 500 times. We compare MAG-GLTrans ($10\times$) with two competitors at different magnification levels ($10\times$, top) and ($20\times$, bottom). \textbf{e-h}, Mean computational time (\textbf{e, g}) of each model and mean size of the WSI files (\textbf{f, h}) in two different tasks.}
	\label{supp-fig3}
\end{figure}

\begin{figure}[!]
	\centering
	\includegraphics[width=.985\linewidth]{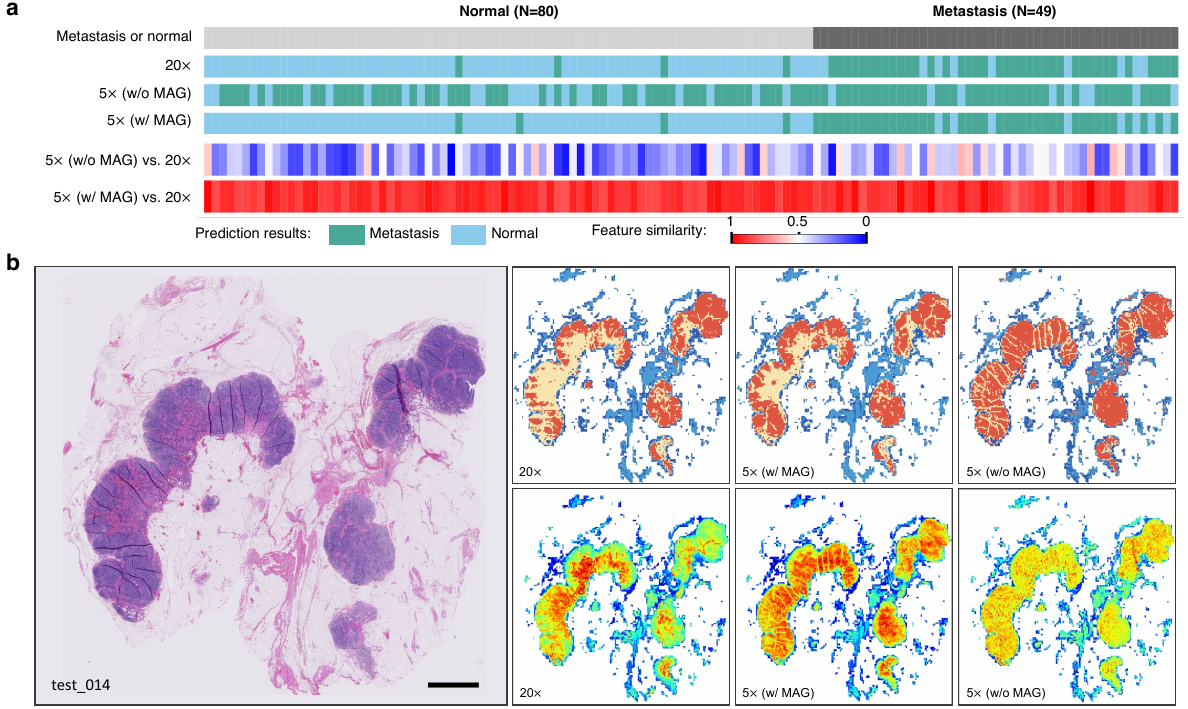}
	\caption{\textbf{Visualization of the features extracted by MAG (CAMELYON16).} \textbf{a}, The prediction results (top) of the test set on the breast cancer lymph node metastasis detection task (CAMELYON16, $n=129$) under different magnification levels. Patients are sorted by the ground truth labels. Heatmaps (bottom) demonstrate the similarity of the features between high magnification level and low magnification level (with and without MAG). We show the feature similarity in slide level, which is the mean of all the patches. Red represents higher similarity. \textbf{b}, From top left to bottom right: a sample case WSI of metastasis, visualization of the unsupervised clustering of the patch-level features (top right) and the heatmap (bottom right) at $20\times$, $5\times$ with MAG and $5\times$ without MAG.}
	\label{supp-fig4}
\end{figure}

\newpage
\bibliography{refs}

\noindent\textbf{Acknowledgements}\\
This work was supported by Noncommunicable Chronic Diseases-National Science and Technology Major Project (No. 2024ZD0531100),
Guangdong Provincial Key Laboratory of Artificial Intelligence in Medical Image Analysis and Application (No. 2022B1212010011),
Regional Innovation and Development Joint Fund of National Natural Science Foundation of China (No. U22A20345), 
Natural Science Foundation for Distinguished Young Scholars of Guangdong Province (No. 2023B1515020043) and
National Natural Science Foundation of China (No. 82372044, 82472062 and 82272084).\\

\noindent\textbf{Author contributions}\\
C.H., B.Z. and Z.L. designed the approach, experiments and drafted the manuscript. B.Z. wrote the code. J.L., and S.L. collected and processed the dataset. L.W. developed the interactive demo. C.H., B.Z., T.D., C.L., X.G., H.W. and Z.S. provided statistical analysis and interpretation of the data. C.H., C.L., Z.S., and Z.L. coordinated and supervised the whole work. All authors were involved in critical revisions of the manuscript, and have read and approved the final version.\\

\noindent\textbf{Declaration of interests}\\
The authors declare no competing interests. \\


\end{document}